\documentclass[lettersize,journal]{IEEEtran}
\usepackage{amsmath,amsfonts}
\usepackage{algorithmic}
\usepackage{algorithm}
\usepackage{array}
\usepackage[caption=false,font=normalsize,labelfont=sf,textfont=sf]{subfig}
\usepackage{textcomp}
\usepackage{stfloats}
\usepackage{url}
\usepackage{verbatim}
\usepackage{graphicx}
\usepackage{cite}
\usepackage{multirow}
\usepackage{pdflscape}
\usepackage{threeparttable}
\usepackage{booktabs}

\usepackage{tabularx}
\usepackage{array}
\newcolumntype{L}[1]{>{\raggedright\arraybackslash}p{#1}}
\newcolumntype{Y}{>{\raggedright\arraybackslash}X}

\begin{document}

\title{READ: A Retrieval-Alignment Diffusion Framework for Structure-based Drug Design}

\author{Dong~Xu,
        Zhangfan~Yang,
        Junchuang Cai,
        Sisi Yuan,
        Zexuan~Zhu,
        Jianqiang~Li,
        and~Junkai~Ji
\thanks{Dong Xu, Junchuang Cai, Sisi Yuan, Zexuan Zhu, Jianqiang Li, and Junkai Ji are with the School of Artificial Intelligence, Shenzhen University, Shenzhen 518060, China.}
\thanks{Zhangfan Yang is with the School of Computer Science, University of Nottingham Ningbo, Ningbo 315100, China.}
\thanks{Corresponding author: Junkai Ji (e-mail: jijunkai@szu.edu.cn.)}
\thanks{\copyright~2026 IEEE. Personal use of this material is permitted. Permission from IEEE must be obtained for all other uses, in any current or future media, including reprinting/republishing this material for advertising or promotional purposes, creating new collective works, for resale or redistribution to servers or lists, or reuse of any copyrighted component of this work in other works. This is the accepted version of an article published in \emph{IEEE Transactions on Computational Biology and Bioinformatics}, doi: 10.1109/TCBBIO.2026.3721445.}
}

\markboth{IEEE Transactions on Computational Biology and Bioinformatics}
{Xu \MakeLowercase{\textit{et al.}}: READ: A Retrieval-Alignment Diffusion for Structure-based Drug Design}

\maketitle

\begin{abstract}
Structure-based drug design (SBDD) models are central to modern pharmaceutical research, enabling the rational exploration of protein–ligand interactions at atomic resolution. However, most existing approaches frame molecular generation as an isolated optimization or a one-to-one matching task, overlooking the shared binding patterns and intrinsic similarities among protein–ligand complexes. This fragmented perspective constrains their ability to capture the fundamental principles governing molecular recognition and binding specificity. Moreover, the limited availability of high-quality experimental data further hampers model generalization and real-world applicability. To address these challenges, we present READ, a retrieval–alignment molecular generation framework that conditions the generative process on small molecules targeting homologous proteins. Retrieved ligands are aligned with a diffusion model across multiple representational spaces and integrated as conditional guidance throughout successive stages of generation. Under a standardized docking-based evaluation protocol, READ achieves consistently strong performance against state-of-the-art SBDD methods. More importantly, it introduces a retrieval-alignment paradigm for structure-based molecular generation, offering a practical framework for early-stage computational hit generation while leaving prospective experimental validation as future work.
\end{abstract}


\section{Introduction}
\IEEEPARstart{D}{rug} discovery is an inherently time-consuming and capital-intensive process, characterized by an exceedingly low success rate from initial compound identification to clinical approval. Traditional in silico techniques, such as virtual screening (VS) based on molecular docking, have been developed to accelerate this pipeline by computationally evaluating the binding affinities between candidate molecules and target proteins\cite{paggi2024art}. However, the efficiency and reliability of these approaches deteriorate as the screening libraries expand exponentially, often leading to high false-positive rates and computational costs\cite{hu2023deep, li2025decoding}. Recently, structure-based molecular generation has emerged as a promising alternative paradigm. Instead of exhaustively screening pre-existing molecular databases, these methods directly generate candidate ligands conditioned on the three-dimensional structure of a target protein. This end-to-end generation framework not only bypasses the inefficiencies inherent in large-scale virtual screening, but also enables optimized binding characteristics and chemical diversity \cite{du2024machine}.

Autoregressive approaches have played a central role in structure-based molecule generation, as they enable sequential modeling of atomic and structural dependencies within the protein binding pocket. These methods typically construct molecular structures atom by atom or fragment by fragment, conditioning each generation step on both the pocket context and previously generated components. Pocket2Mol learns context-specific distributions of atom and bond types for each spatial position inside the pocket, generating candidate ligands by iteratively sampling from these learned conditional distributions\cite{pocket2mol}. Lingo3DMol extends this paradigm by integrating language modeling with geometric deep learning, thereby jointly capturing molecular topology and 3D spatial organization in a pocket-conditioned generative process\cite{lingo3dmol}. To better capture hierarchical dependencies, ResGen formulates the generation task as a two-level autoregression, where a global module learns protein–ligand interaction patterns, and a local atomic module refines fine-grained geometric distributions\cite{resgen}. FragGen further advances autoregressive generation by decomposing molecular geometry into multiple sets of geometric variables and adopting a hybrid autoregressive strategy, which enhances both the geometric fidelity and synthesizability of generated molecules\cite{fraggen}. Overall, autoregressive models offer a flexible and interpretable framework for molecule generation; however, they often suffer from error accumulation during the stepwise construction process. Moreover, the independent stepwise construction process is difficult to capture collective molecular interactions and global dependency patterns within the generated structures.

Recent progress in geometric deep learning techniques, including diffusion and flow models, can overcome these limitations by treating molecular generation as a holistic process, enabling globally consistent structure formation and naturally incorporating molecular symmetries. For instance, DiffLinker introduced an E(3)-equivariant conditional diffusion framework that autonomously determines attachment points to connect arbitrary molecular fragments\cite{difflinker}. Extending this principle to structure-based drug design, DiffSBDD generates novel ligands conditioned on protein pockets while respecting translational, rotational, and permutational symmetries\cite{diffsbdd}. DiffBP further incorporates target proteins as contextual constraints to model full-atom molecular 3D structures\cite{diffbp}. To enhance drug-likeness and synthesizability, DecompDiff decomposes ligands into scaffold and arm components, coupling atom- and bond-level diffusion processes\cite{decompdiff}. Meanwhile, flow-matching approaches such as EquiFM and ET-Flow achieve stable and efficient geometric generation by leveraging equivariance and harmonic priors\cite{song2023equivariant, hassan2024flow}. However, due to the limited availability of experimentally resolved protein–ligand complexes, current geometric generative models cannot reliably produce generalizable and biologically plausible molecules for specific target proteins\cite{alakhdar2024diffusion, guo2024diffusion}.

Retrieval-augmented generation (RAG) techniques have been widely used in natural language processing and computer vision, which leverage a pretrained retriever to extract relevant knowledge from external databases to enhance generation\cite{lewis2020retrieval, alayrac2022flamingo}. Protein–ligand interactions often exhibit high structural similarity, homologous proteins tend to bind to the same or chemically related ligands, while structurally similar small molecules can also target the same proteins\cite{elijovsius2025zero}. Leveraging these structural priors through RAG methods offers a promising solution for geometric deep learning methods to alleviate the scarcity of training data by incorporating external structural knowledge from homologous proteins and known ligands. Following this perspective, a retrieval-alignment diffusion framework named READ is proposed in this study, which integrates RAG with E(3)-equivariant generative modeling to enable biologically plausible and generalizable structure-based molecular design.

READ retrieves ligands from homologous proteins and aligns their multi-level information with a diffusion model during both training and inference. Given a target protein, READ first retrieves homologous proteins based on sequence and structural similarity, employing TM-align\cite{TMalign} and DaliLite\cite{DaliLite} for multiple sequence and structural alignments, respectively. A molecular encoder is pretrained on the ultra-large-scale compound database ZINC20 using contrastive learning\cite{zinc20, MolCLR}, extracting multi-level features of bound ligands from homologous proteins, including atom-level, fragment-level, and graph-level representations. The learned representations inherently capture docking-relevant information, reflecting interactions between the protein family and their bound ligands, and serve as conditional inputs to guide ligand generation. Through a projection MLP, representations from the encoder’s high-level, intermediate, and low-level layers are separately aligned with the corresponding reverse layers of the diffusion model. During inference, these representations are gradually injected in a stepwise manner, as the conditional information required during the denoising process increases over time. In addition, the denoising model treats protein–protein edges as rigid, whereas protein–ligand and ligand–ligand edges are treated as flexible. This design preserves the overall protein scaffold while allowing sufficient flexibility for ligand binding and conformational adjustments, enabling accurate modeling of pocket–ligand interactions.

The main contributions of this study can be summarized as follows.

(i) Retrieval-based ligand conditioning: READ introduces a novel paradigm that leverages retrieved ligands as informative priors to guide pocket-conditioned molecular generation.

(ii) Layer-wise alignment and stepwise guidance: READ employs multi-level representation alignment and stage-wise diffusion conditioning to preserve rational interaction patterns and maintain structural topology consistency.

(iii) Docking-based empirical performance: Under a unified CrossDocked2020 evaluation protocol, READ achieves strong docking-based scores, improvement rates, and pocket-level stability compared with existing generative models.

The remainder of this paper is organized as follows. Section II introduces the problem setup, notation, diffusion processes, and the ligand-size scale prior. Section III presents the proposed READ framework. Section IV describes the experimental setup. Section V reports the experimental results. Section VI discusses implications, limitations, and scalability. Section VII concludes the paper.

\begin{figure*}[htbp] 
\centering \includegraphics[width=0.81\textwidth]{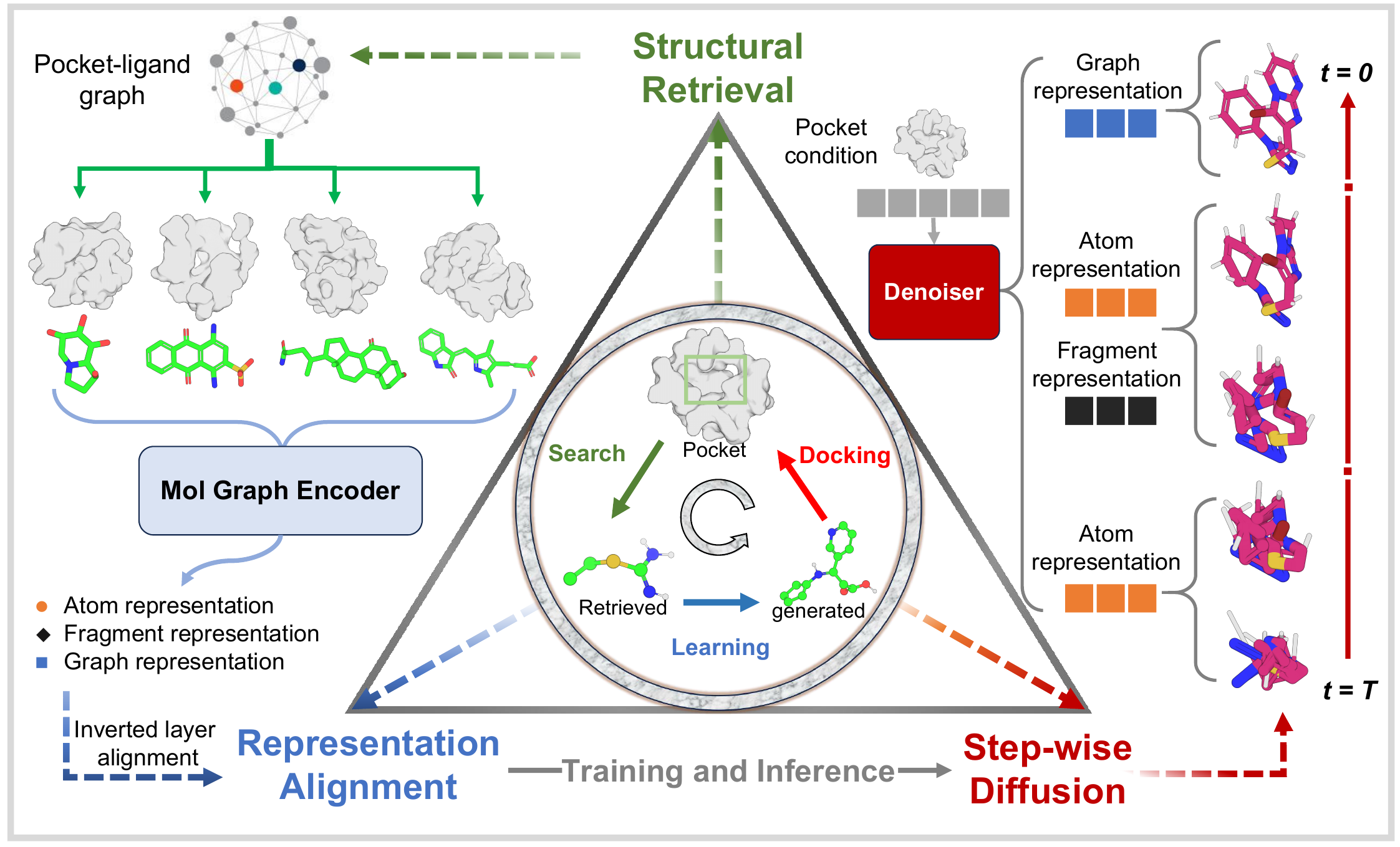} 
\vspace{-2mm}
\caption{Overview of READ. A pocket–ligand graph is used to support structural retrieval, representation alignment, and stepwise diffusion. Retrieved ligands are encoded as ligand side priors by a molecular graph encoder and are matched to the target pocket. These priors are coupled to the denoiser through representation alignment, and are injected along the diffusion trajectory so that shape signals guide early steps, interaction signals guide middle steps, and atom types guide late steps.}
\vspace{-4mm}
\label{fig:fig1} 
\end{figure*}

\section{PRELIMINARIES}\label{sec:preliminaries}

\subsection{Problem Setup and Notation}\label{sec:problem_setup}

Given a protein pocket $P$, the task is to generate a ligand $X$ with atomic coordinates and element types that are compatible with the binding site. A protein pocket is written as
\begin{equation}
P = (R_P, Z_P),
\end{equation}
where $R_P \in \mathbb{R}^{n_P \times 3}$ are receptor atom coordinates and $Z_P \in \mathcal{A}_P^{n_P}$ are receptor element types for $n_P$ atoms. During training, each pocket is cropped by a sphere of radius $10$~\AA\ centered at the ligand center of mass. The cropped pocket is then recentered to remove global translation, and to define a ligand centered local coordinate system.

The ligand is denoted by
\begin{equation}
X = V = \{(r_i, a_i)\}_{i=1}^N,
\end{equation}
where $r_i \in \mathbb{R}^3$ and $a_i \in \mathcal{A}$ are the coordinate and element type of the $i$th atom, and
\begin{equation}
\mathcal{A} = \{\mathrm{C}, \mathrm{N}, \mathrm{O}, \mathrm{F}, \mathrm{P}, \mathrm{S}, \mathrm{Cl}, \mathrm{Br}, \mathrm{I}\}, \quad N \le 35.
\end{equation}
The bound $N \le 35$ matches the heavy atom range considered in the benchmark dataset and is imposed for all models. All ligand coordinates are stacked into a vector $x \in \mathbb{R}^{3N}$. Atom types are embedded as a relaxed categorical variable using a logistic normal representation and stacked into a vector $y$. This logistic normal relaxation is referred to as RelaxCat. It approximates discrete atom types by continuous logits whose softmax lies on the probability simplex. This representation supports type diffusion and gradient-based optimization. We use RelaxCat because it keeps atom-type variables in a continuous and differentiable space, which makes them naturally compatible with coordinate diffusion and gradient-based inference guidance. In contrast, Gumbel-Softmax mainly provides a differentiable sampling approximation and can introduce temperature-sensitive stochasticity, while fully discrete diffusion or discrete score matching would require a separate categorical transition process and make joint coordinate--type guidance less direct. RelaxCat therefore provides a practical compromise for jointly denoising continuous coordinates and categorical atom types. The supervision targets of the generative model are $(x_0, y_0)$, the coordinates and relaxed types of the native ligand, and $a_0$ denotes the discrete atom types corresponding to $y_0$.

The conditional distribution
\begin{equation}
p_{\theta}(X \mid P) = p_{\theta}(x, y \mid P)
\end{equation}
is learned from paired pockets and ligands. At inference, only the processed pocket $P$ is given and a ligand sample $\hat{X}$ is drawn from the learned conditional distribution.

\subsection{Diffusion Processes and Basic Objective}\label{sec:diffusion_process}

Ligand generation is formulated as a discrete time diffusion process on both coordinates and relaxed atom types. The forward processes gradually corrupt $(x_0, y_0)$ into noise. A denoiser is trained to invert this corruption conditioned on the pocket.

For coordinates, a Gaussian forward process with $T$ timesteps is defined as
\begin{equation}
q(x_t \mid x_{t-1}) = \mathcal{N}\bigl(\sqrt{1-\beta^{(x)}_t}\, x_{t-1},\, \beta^{(x)}_t I_{3N}\bigr),
\end{equation}
where $\{\beta^{(x)}_t\}_{t=1}^{T}$ is a variance schedule. A sigmoid schedule is adopted with $\beta^{(x)}_1 = 10^{-7}$, $\beta^{(x)}_T = 2\times 10^{-3}$, and $T = 2000$. The schedule keeps early timesteps close to the data manifold and late timesteps near an isotropic Gaussian distribution.

For the relaxed type variables, the forward process is defined in the RelaxCat space as
\begin{equation}
q(y_t \mid y_{t-1}) = \mathrm{RelaxCat}\bigl(y_{t-1}; \beta^{(y)}_t\bigr),
\end{equation}
with a cosine schedule for $\{\beta^{(y)}_t\}_{t=1}^{T}$ (smoothing $s = 0.01$) and the same number of timesteps. The RelaxCat kernel smoothly interpolates between the original categorical logits and an almost uniform distribution. It serves as a continuous analogue of diffusion on atom types. In the forward direction, the coordinate and type processes are conditionally independent given the pocket. In the reverse direction, they are coupled only through the denoiser.

The denoiser $U_{\theta}$ receives the pocket and noisy ligand variables $(P, x_t, y_t)$ and predicts both coordinate noise and atom type logits. For coordinates, the denoiser predicts an estimate $\varepsilon^{(x)}_{\theta}(P, x_t, y_t, t)$ of the Gaussian noise. For atom types, the denoiser outputs logits that parameterize a categorical distribution over $\mathcal{A}$ under the logistic normal relaxation. The basic diffusion loss is
\begin{equation}
\mathcal{L}_{\mathrm{diff}} =
  w_{\mathrm{pos}}\, \mathbb{E}\bigl[\lVert \varepsilon^{(x)} - \varepsilon^{(x)}_{\theta} \rVert^{2}_{2}\bigr]
+ w_{\mathrm{atom}}\, \mathbb{E}\bigl[\mathrm{CE}(a_{0}, \pi^{\mathrm{atom}}_{\theta})\bigr],
\end{equation}
where $\varepsilon^{(x)}$ denotes Gaussian noise on coordinates, $a_0$ are ground truth atom types, and $\pi^{\mathrm{atom}}_{\theta}$ is the categorical prediction induced by the denoiser. The weights $w_{\mathrm{pos}}$ and $w_{\mathrm{atom}}$ balance coordinate and type supervision. This objective is used in all variants of READ and is later augmented by alignment and guidance terms.

\subsection{Scale Prior on Ligand Size}\label{sec:scale_prior}

In addition to the diffusion processes, a scale prior on ligand size is introduced. The prior is defined from a pocket level statistic computed on the cropped pocket coordinates $R_P$. Let
\begin{equation}
d_{ij} = \lVert r^{(P)}_i - r^{(P)}_j \rVert
\end{equation}
denote pairwise Euclidean distances between pocket atoms with coordinates $r^{(P)}_i$. Distances are sorted in descending order, and the median of the ten largest values is taken as
\begin{equation}
S(P) = \operatorname{median}_{\text{top }10}\{ d_{ij} \}.
\end{equation}
This statistic provides a stable estimate of the linear scale of the binding site and is relatively insensitive to local perturbations.

The statistic $S(P)$ is discretized into bins $\{b_1,\dots,b_K\}$ obtained from equal frequency quantiles on training pockets. For each bin $i$, a categorical distribution $p_i(N)$ over heavy atom counts $N \in \{1,\dots,35\}$ is estimated from the ligands assigned to that bin. The empirical histograms are smoothed and normalized to avoid zero probabilities. The bin boundaries and per bin tables are precomputed on the training set and remain fixed during training and evaluation.

At inference time, $S(P)$ is computed for a test pocket and the corresponding bin index $i$ is identified. A heavy atom count is then sampled as $N \sim p_i(N)$. The sampled count $N$ specifies the number of active ligand atoms in the generative process. Existence masks over atom indices are used so that the effective atom count reaches $N$ early in the denoising trajectory. In later steps, the model mainly refines coordinates and poses after the ligand topology has stabilized. This mechanism decouples coarse control of ligand size from fine scale conformational and chemical refinement.
This scale prior is used only as an initialization-level coarse size controller. It samples a plausible heavy-atom count from empirical pocket-scale bins and then leaves the denoiser to refine geometry, pose, and chemistry. It is not an additional atom-count prediction head, nor is it a direct atom-count supervision loss inside the denoiser. This distinguishes it from explicit atom-count priors used in some pocket-conditioned diffusion baselines. Because the bins are constructed by equal-frequency quantiles over training pockets, each bin has comparable empirical support, which reduces sensitivity to manually chosen absolute scale thresholds.

\section{Methods}\label{sec:methods}
The READ framework combines structural retrieval, representation alignment, and stepwise diffusion for pocket conditioned ligand generation, as summarized in Fig.~\ref{fig:fig1}.

\begin{figure*}[htbp] 
\centering \includegraphics[width=0.86\textwidth]{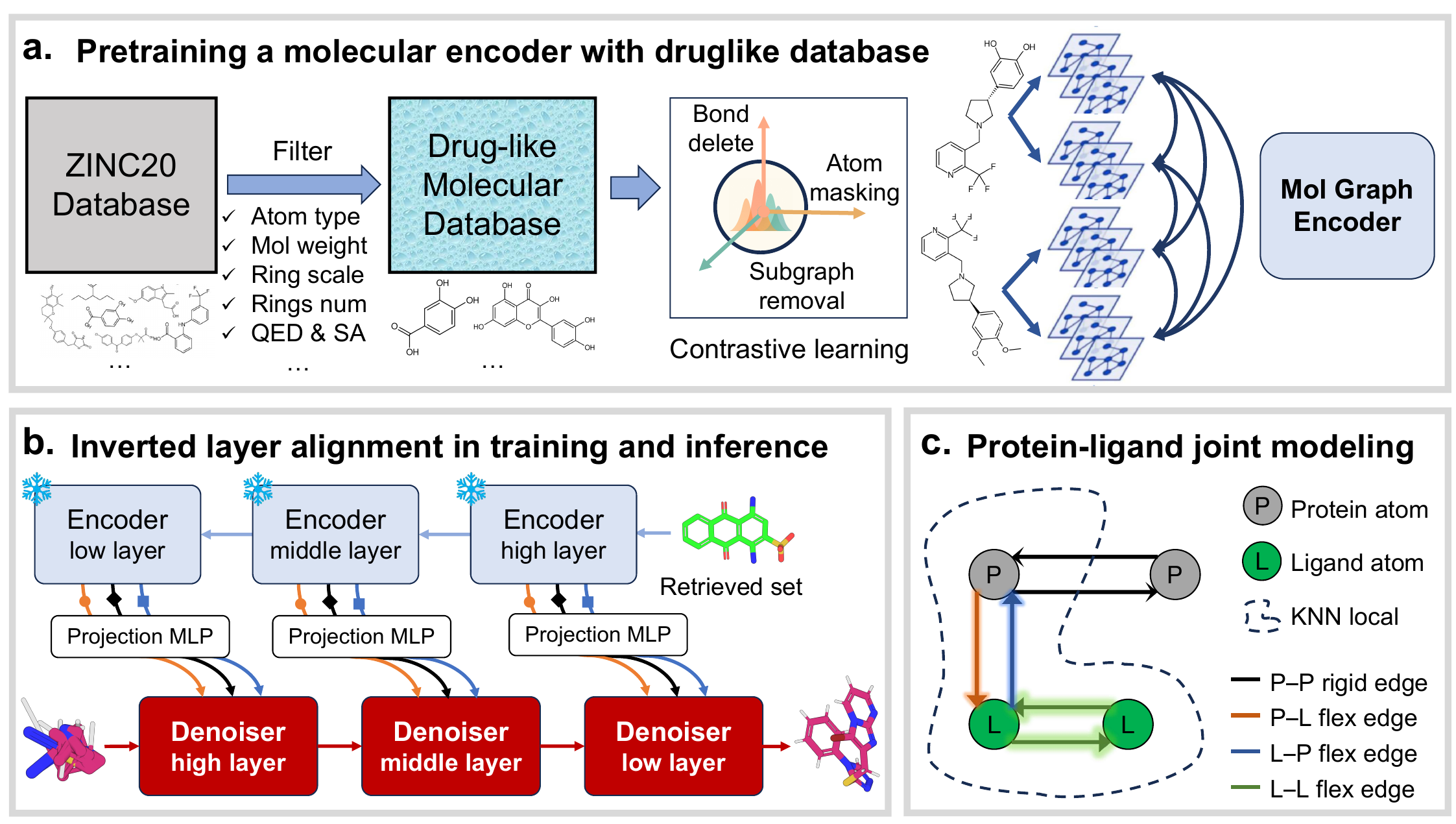} 
\vspace{-2mm}
\caption{Components of READ. (a) Pretraining of a molecular graph encoder on a filtered drug-like subset of ZINC20 with contrastive augmentations, which produces the embedding space used for retrieval and guidance. (b) Inverted layer alignment during training and inference, where low, middle, and high encoder layers are projected to late, middle, and early denoiser layers, respectively. (c) Joint protein–ligand graph used by the denoiser, with rigid edges among protein atoms and flexible edges for protein–ligand and ligand–ligand neighborhoods.}
\vspace{-4mm}
\label{fig:fig2} 
\end{figure*}

\subsection{Graph Representation of Pocket and Ligand}\label{sec:graph_representation}

READ uses pocket and ligand graphs defined in the same ligand-centered coordinate system for conditioning and supervision. For each cropped pocket $P=(R_P,Z_P)$ defined in Section~\ref{sec:problem_setup}, a pocket graph $G_P$ is constructed with one node for each pocket atom. Node features contain element type encodings and three dimensional coordinates. Edges connect local neighboring atoms. Edge features store relative position vectors and geometric descriptors for the equivariant message passing network. During training, isotropic Gaussian noise with standard deviation $\sigma = 0.1$~\AA\ is added to pocket atom coordinates to reflect experimental and docking uncertainty. This perturbation is not applied at inference.

The ligand $X$ is represented as an atom graph. Each node stores an element type $a_i$ and a coordinate $r_i$ taken from the protein ligand complex. Coordinates are expressed in the same ligand centered coordinate system as $G_P$. Edges encode chemical bonds and local neighborhoods with geometric features. Pocket and ligand graphs are merged into a joint graph with binary masks. These masks distinguish receptor and ligand atoms and separate type and coordinate channels. This joint graph supports simultaneous supervision of atom types and coordinates during training. At de novo inference, only the pocket graph is provided. The ligand part is initialized from noise according to the diffusion processes in Section~\ref{sec:diffusion_process}. The denoiser $U_{\theta}$ then operates on the resulting joint graph. The resulting joint protein–ligand graph with rigid and flexible edges is depicted in Fig.~\ref{fig:fig2}(c).

\subsection{Molecular Encoder Pretraining}\label{sec:encoder_pretraining}

The molecular encoder is pretrained on a large corpus of two dimensional molecular graphs before being integrated into READ. A subset of approximately $8\times 10^{6}$ molecules is sampled from ZINC20~\cite{zinc20} under the following filters: $300 < \mathrm{Mw} < 600$, quantitative estimate of drug-likeness greater than $0.6$, at most five rings in the smallest set of smallest rings, no macrocycles with at least eight members, elements restricted to ${\mathrm{C}, \mathrm{N}, \mathrm{O}, \mathrm{F}, \mathrm{P}, \mathrm{S}, \mathrm{Cl}, \mathrm{Br}, \mathrm{I}}$, and synthetic accessibility score $\mathrm{SA} \le 6.0$. Two dimensional molecular graphs parsed from SMILES strings are used as inputs. They are treated independently of protein structures. This pretraining stage yields an embedding space that captures drug-like chemistry. It is later reused for retrieval, distilled priors, and alignment, as illustrated in Fig.~\ref{fig:fig2}(a).

Pretraining uses a contrastive objective with graph augmentations. For each molecule, two augmented views are generated by perturbing the graph, for example by masking atoms or dropping edges. The encoder is optimized with an InfoNCE loss over the augmented pairs with a temperature scaled softmax~\cite{MolCLR}. Representations of the same molecule are pulled together, and representations of different molecules are pushed apart. The conditional diffusion model is then trained on protein–ligand complexes while the pretrained encoder is either finetuned or partially frozen. The specific settings are described in Section~\ref{sec:training}.

\subsection{Pocket and Ligand Atlas, Retrieval, and Distillation}\label{sec:retrieval_distillation}

An atlas $\mathcal{A}=\{(P_k,L_k)\}$ of pockets and their validated ligands or high quality docked surrogates is constructed from protein–ligand complexes~\cite{edge2024local,irdiff}. The atlas aggregates nonredundant complexes with sequence clustering at $95\%$ identity and standardized ligands. Complexes corresponding to held out test targets are excluded from the atlas to avoid information leakage. For a target pocket $P$, sequence and structure similarities to atlas entries are first computed. A neighbor set is then formed as
\begin{equation}
\mathcal{N}_K(P) = \mathrm{TopK}_k\big(s(P,P_k)\big),
\end{equation}
where $s(P,P_k)$ denotes the combined similarity between $P$ and $P_k$. Unless stated otherwise, $K=40$ neighbors are retrieved and $m=4$ ligands per neighbor are retained after filtering. This procedure yields an aligned ligand set associated with $P$. Sequence and structure similarities are $z$ normalized across the atlas and combined as
\begin{equation}
s(P,P_k) = \alpha s_{\mathrm{seq}}(P,P_k) + (1-\alpha)s_{\mathrm{str}}(P,P_k),
\end{equation}
with $\alpha = 0.5$. Each neighbor complex is rigidly aligned to the target pocket before ligand distillation.

From the aligned ligands $\{L_k\}$, three molecular representations usable by the diffusion model are extracted,
\begin{equation}
\mathcal{R}(P) = \big\{\mathbf{V}_{\mathrm{occ}},\, \mathbf{I}_{\mathrm{pat}},\, \mathbf{C}\big\}.
\end{equation}
Here $\mathbf{V}_{\mathrm{occ}}$ denotes a voxelized occupancy field and is distinct from the scalar scale statistic $S(P)$ defined in Section~\ref{sec:scale_prior}.

A continuous occupancy field is first constructed by kernel density estimation,
\begin{equation}
F_{\mathrm{occ}}(\mathbf{x}) = \sum_{i}\exp\!\left(-\frac{\|\mathbf{x}-\mathbf{r}^{(L)}_i\|^2}{\tau^2}\right),
\end{equation}
where $\mathbf{r}^{(L)}_i$ are ligand atom coordinates from the retrieved set. The voxel resolution is $1.25$~\AA{} and the kernel bandwidth is $\tau=1.5$~\AA{}. Optional channels such as accessibility, hydrophobic versus polar partitions, and electrostatics can be added to $\mathbf{V}_{\mathrm{occ}}$. Unless stated otherwise, only the occupancy channel is used in the main experiments.

Interaction patterns are represented by a bipartite graph $G_{\mathrm{pat}}=(V_P,V_L,E_{\mathrm{pat}})$ between candidate pocket sites $V_P$ and ligand pharmacophore sites $V_L$. Edge weights combine geometry and type compatibility,
\begin{equation}
w_{p\ell} = \rho(d_{p\ell})\,\gamma(\angle_{p\ell})\,\chi(\text{type-compatible}),
\end{equation}
with a $2.5$–$4.0$~\AA{} distance window and standard angular tolerances for hydrogen bonds and $\pi$ contacts.

Atom preferences are captured by estimating conditional distributions $\Pr(a\mid \mathrm{env})$ over element types. In the local environment aligned to $P$ at distance shells $3$, $6$, and $9$~\AA{}, empirical distributions over atom types are estimated by kernel density estimation. These distributions define $\mathbf{C}$. After this step, pocket conditioning is converted to molecular-side signals $\mathbf{V}_{\mathrm{occ}},\mathbf{I}_{\mathrm{pat}},\mathbf{C}$. Subsequent guidance interacts only with $\mathcal{R}(P)$ and does not use raw protein features.

\subsection{Inverted Layer Wise Alignment from the Encoder to the Denoiser}\label{sec:alignment}

The molecular encoder $E_{\mathrm{pre}}$ from the pretraining stage exposes intermediate activations $\{\mathbf{h}^{\mathrm{pre}}_\ell\}_{\ell=1}^{L}$. The denoiser $U_{\theta}$ from Section~\ref{sec:diffusion_process} exposes hidden states $\{\mathbf{h}^{\mathrm{den}}_m\}_{m=1}^{M}$. Projection heads $\phi_{\mathrm{occ}},\phi_{\mathrm{pat}},\phi_C$ map the distilled priors into encoder aligned tokens
\begin{equation}
\begin{aligned}
\mathbf{z}_{\mathrm{occ}} &= \phi_{\mathrm{occ}}(\mathbf{S}) \approx \mathbf{h}^{\mathrm{pre}}_{L},\\
\mathbf{z}_{\mathrm{pat}} &= \phi_{\mathrm{pat}}(\mathbf{I}_{\mathrm{pat}}) \approx \mathbf{h}^{\mathrm{pre}}_{\lfloor L/2\rfloor},\\
\mathbf{z}_C &= \phi_C(\mathbf{C}) \approx \mathbf{h}^{\mathrm{pre}}_{1}.
\end{aligned}
\end{equation}
This design differs from simple feature concatenation or FiLM-style conditioning. The retrieved priors are not injected as a single global conditioning vector. Instead, they are matched to denoiser states at different semantic depths, so that coarse occupancy, interaction patterns, and atom-level preferences influence the stages of denoising where they are most relevant. This alignment can be viewed as regularizing the denoiser representation space before inference-time guidance is applied, thereby making the subsequent scheduled guidance more stable and effective.
\begin{equation}
\begin{aligned}
\tilde{\mathbf{h}}^{\mathrm{den}}_{m} &= \mathrm{Attn}\!\big(\mathbf{h}^{\mathrm{den}}_{m}W_Q,\, \mathbf{z}_{\star(m)}W_K,\, \mathbf{z}_{\star(m)}W_V\big),\\
\widehat{\mathbf{h}}^{\mathrm{den}}_{m} &= (1-\alpha_m(t))\,\mathbf{h}^{\mathrm{den}}_{m}+\alpha_m(t)\,\tilde{\mathbf{h}}^{\mathrm{den}}_{m},
\end{aligned}
\end{equation}
where $\star(1)=S$, $\star(\lfloor M/2\rfloor)=\mathrm{pat}$, and $\star(M)=C$, and $W_Q,W_K,W_V$ are learned projections. Cross attention is applied only at these depths. For other $m$, the gating weight $\alpha_m(t)$ is set to zero and $\widehat{\mathbf{h}}^{\mathrm{den}}_{m}$ reduces to $\mathbf{h}^{\mathrm{den}}_{m}$. Deeper encoder features capture shape and volume, middle layers capture interaction patterns, and shallow layers capture atomic detail. The denoiser requires coarse geometry early and fine chemistry late, so the semantic depths are aligned accordingly. The mapping from encoder layers to denoiser layers and the resulting information flow are shown schematically in Fig.~\ref{fig:fig2}(b).

The encoder and denoiser are coupled via an inverted layer wise alignment objective. At diffusion timestep $t$, let $h_{m,t}$ denotes the $m$th denoiser hidden state from $U_{\theta}$. Let $P_m(\cdot)$ project the $m$th denoiser hidden state into the encoder space. Let $z^{\mathrm{target}}_{m}$ be the encoder feature derived from pocket conditioned priors at the matched semantic depth~\cite{yu2024representation}. The alignment loss is
\begin{equation}
\begin{aligned}
\hat{h}_{m,t} &= \mathrm{LayerNorm}\big(P_m(h_{m,t})\big),\\
\mathcal{L}_{\mathrm{align}} &= \sum_{t}\sum_{m}\gamma_m(t)\,d\!\left(\hat{h}_{m,t},\,z^{\mathrm{target}}_{m}\right),
\end{aligned}
\end{equation}
where $d$ is $1-\cos$ similarity or $\ell_2$ in ablations and $\gamma_m(t)$ schedules early, middle, and late emphasis. The denoiser objective is
\begin{equation}
\mathcal{L}_{\mathrm{denoiser}} = \mathcal{L}_{\mathrm{diff}} + \lambda_{\mathrm{align}}\mathcal{L}_{\mathrm{align}},
\end{equation}
where $\mathcal{L}_{\mathrm{diff}}$ is the basic diffusion loss defined in Section~\ref{sec:diffusion_process}.

\subsection{Guidance at Inference}\label{sec:guidance}

During inference, guidance is applied at each timestep using the retrieved priors. At timestep $t$, the trained encoder is reused to obtain $\mathbf{z}_{\mathrm{occ}},\mathbf{z}_{\mathrm{pat}},\mathbf{z}_C$ from retrieved ligands. Denoising is guided by scheduled weights. The coordinate noise prediction $\boldsymbol{\varepsilon}^{(x)}_\theta$ defined in Section~\ref{sec:diffusion_process} is corrected by gradients from three energies:
\begin{equation}
\begin{aligned}
\hat{\boldsymbol{\varepsilon}}^{(x)}_\theta
&= \boldsymbol{\varepsilon}^{(x)}_\theta
 - \lambda_{\mathrm{occ}}(t)\,\nabla_{\mathbf{x}_t}\mathcal{E}_{\mathrm{occ}}(\mathbf{x}_t;\mathbf{z}_{\mathrm{occ}}) \\
&\quad - \lambda_{\mathrm{pat}}(t)\,\nabla_{\mathbf{x}_t,\mathbf{y}_t}
          \mathcal{E}_{\mathrm{pat}}(\mathbf{x}_t,\mathbf{y}_t;\mathbf{z}_{\mathrm{pat}}) \\
&\quad - \lambda_C(t)\,\nabla_{\mathbf{y}_t}\mathcal{E}_C(\mathbf{y}_t;\mathbf{z}_C).
\end{aligned}
\end{equation}

For the shape and volume term, a ligand occupancy $\hat{F}_S(\mathbf{x})$ is rendered in the current coordinate frame. The energy is defined as $\mathcal{E}_{\mathrm{occ}}(\mathbf{x}_t;\mathbf{z}_{\mathrm{occ}})=|\hat{F}_S-F_{\mathrm{occ}}|_2^2$. Jensen–Shannon divergence is used in ablations.

For the interaction pattern term, the energy over site and pharmacophore pairs is
\begin{equation}
\begin{aligned}
\mathcal{E}_{\mathrm{pat}}(\mathbf{x}_t,\mathbf{y}_t;\mathbf{z}_{\mathrm{pat}})
&= \sum_{(p,\ell)\in E_{\mathrm{pat}}}
    w_{p\ell}\,\big(\|\mathbf{r}_\ell-\mathbf{r}_p\|-\delta_{p\ell}\big)^2 \\
&\quad
  + \sum_{(p,\ell)\in E_{\mathrm{pat}}} \omega_{p\ell}\,\Psi_{\mathrm{angle/dihedral}},
\end{aligned}
\end{equation}
where $\mathbf{r}_p$ and $\mathbf{r}_\ell$ are pocket and ligand coordinates for interaction pairs, $\delta_{p\ell}$ are preferred distances, and $\Psi_{\mathrm{angle/dihedral}}$ penalizes angular and dihedral deviations.

For the atom type term, the atom preference energy is
\begin{equation}
\mathcal{E}_C(\mathbf{y}_t;\mathbf{z}_C) = -\sum_i \log \Pr(a_i\mid \mathrm{env}_i),
\end{equation}
where $\Pr(a_i\mid \mathrm{env}_i)$ is given by the retrieved atom preference table conditioned on the local environment.

Weights follow a simulated annealing schedule that allocates occupancy guidance primarily to early timesteps, interaction guidance to middle timesteps, and atom-type guidance to late timesteps. This schedule follows the coarse-to-fine nature of ligand generation. Global occupancy and volume are determined before fine chemical identities become reliable; interaction patterns are most useful once a coarse pose has formed; and atom-type preferences are most informative near convergence, when local environments are sufficiently stable. With total timesteps $T$, indicator $\mathbf{1}_{\{\cdot\}}$, and $(u)_+=\max(u,0)$,
\begin{equation}
\begin{aligned}
\lambda_{\mathrm{occ}}(t)&=\lambda_S^{\max}\cos^2\!\Big(\frac{\pi t}{2t_S}\Big)\,\mathbf{1}_{\{t\le t_S\}},\\
\lambda_{\mathrm{pat}}(t)&=\lambda_{\mathrm{pat}}^{\max}\sin\!\Big(\frac{\pi (t-t_{\mathrm{pat}}^-)}{t_{\mathrm{pat}}^+-t_{\mathrm{pat}}^-}\Big)_+,\\
\lambda_C(t)&=\lambda_C^{\max}\sin^2\!\Big(\frac{\pi (t-t_C^-)}{2(T-t_C^-)}\Big)_+,
\end{aligned}
\end{equation}
where $t_S$, $t_{\mathrm{pat}}^-$, $t_{\mathrm{pat}}^+$, and $t_C^-$ control the active windows of the three guidance terms. Unless stated otherwise, $t_{\mathrm{pat}}^-=0.30T$, $t_S=0.35T$, $t_{\mathrm{pat}}^+=0.70T$, $t_C^-=0.55T$, and $t_{\mathrm{freeze}}=0.60T$ are used. With these settings, shape guidance is concentrated in the early part of the trajectory, interaction signals span the middle region, and atom type guidance increases toward the end.

The scale prior $p(N\mid P)$ induced from the statistic $S(P)$ in Section~\ref{sec:scale_prior} is used at the start of inference to sample the heavy atom count $N$. Existence masks over atom indices are then applied so that the number of active ligand atoms increases during early timesteps until it reaches $N$. After that point, the mask remains fixed. Inactive atoms do not contribute to the loss or to updates. This mechanism enforces the target atom count while still allowing early growth of the ligand. To reduce late discrete changes, a topology freeze is introduced. At $t\ge t_{\mathrm{freeze}}$, atom types are held fixed and only continuous variables, namely coordinates and relative poses, are updated. Ablations for this mechanism are reported in the results section.

\subsection{Training Procedure and Optimization}\label{sec:training}

Training proceeds in two stages. In the first stage, the molecular encoder is pretrained on the two dimensional corpus described above with a contrastive objective (InfoNCE)~\cite{MolCLR}. Graph augmentations and a temperature scaled softmax are used. In the second stage, the denoiser, the projection heads, and the encoder are jointly optimized on protein-ligand complexes with the denoiser loss $\mathcal{L}_{\mathrm{denoiser}}$ defined in this section.

Optimization uses Adam with a learning rate of $5\times 10^{-4}$, weight decay $0$, $\beta_1=0.95$, and $\beta_2=0.999$. A plateau scheduler reduces the learning rate by a factor of $0.6$ with patience $10$ and a floor of $1\times 10^{-6}$. Training runs for $2{,}500{,}000$ iterations. Batch size is $64$ unless stated otherwise. The diffusion schedules for coordinates and atom types follow the settings described in Section~\ref{sec:diffusion_process}.

\section{EXPERIMENTAL SETUP}

\subsection{Dataset and Preprocessing}
\label{sec:dataset}

Experiments are conducted on the CrossDocked2020 dataset~\cite{crossdocked2020}, a benchmark commonly used in recent work on structure-based molecular generation. Data partitioning follows the training, validation, and test splits established by LiGAN~\cite{LiGAN} and adopted by 3DSBDD~\cite{3DSBDD}. This setting ensures a reproducible comparison and a direct alignment with prior work. With this pocket definition and dataset split, the resulting benchmark comprises $99{,}900$ protein–ligand pairs for training and $100$ pairs each for validation and testing.

For each protein–ligand complex, the binding pocket is defined as all protein atoms within a sphere of radius $10$~\AA\ centered at the ligand center of mass. This definition matches the problem setup in Section~\ref{sec:problem_setup}. To eliminate confounding effects from preprocessing, this pocket definition and the protein processing pipeline are applied consistently to all baselines and to READ. For docking with AutoDock Vina~\cite{Vina}, the same spherical pocket is used as the search space rather than the entire protein. The default cubic search box of AutoDock Vina is therefore not used, since it is inconsistent with the spherical pocket definition commonly adopted in structure-based molecular generation benchmarks.

\subsection{Multidimensional Evaluation Metrics}
\label{sec:metrics}

For each test pocket, $100$ candidate ligands are generated by every method. The generated structures undergo preliminary filtering with Open Babel~\cite{openbabel} and RDKit~\cite{rdkit} to ensure valence and bond integrity. Structures with physically implausible bond lengths or angles are discarded. To ensure a fair comparison across methods, a unified evaluation pipeline is adopted. Docking is performed with a consistent AutoDock Vina configuration, interactions are analyzed with PLIP~\cite{plip}, and energy minimization uses a fixed force field and convergence criterion~\cite{Vina,MMFF94s}.

Generation quality is assessed from multiple aspects with a primary focus on ligand–protein binding. First, the docking score for each generated ligand is computed. The proportion of generated molecules that outperform the reference ligand score is then measured. Second, binding interfaces are characterized by ligand–protein interaction profiles and by the frequency of steric clashes with the pocket. Beyond the binding interface, intrinsic molecular quality is evaluated by comparing distributions of chemical properties, such as atom types, rings, and functional groups, to those of the reference set. It is also evaluated by checking the correctness of internal static geometry. Finally, model stability is quantified across the diverse set of pockets by pocket level binding gain and ligand efficiency statistics. The analysis prioritizes pocket binding effectiveness, with secondary emphasis on distributional similarity and geometric fidelity.

{Improvement Percentage (IMP).}
For each pocket $p$, let $E^{\mathrm{ref}}_p$ be the Vina energy of the reference ligand and $\{E_{p,j}\}_{j=1}^M$ the energies of $M$ generated ligands.
The per-pocket improvement percentage is
\begin{equation}
\mathrm{IMP}_p = \frac{1}{M}\sum_{j=1}^M \mathbf{1}\!\left(E_{p,j} < E^{\mathrm{ref}}_p\right),
\end{equation}
and the final IMP is
\begin{equation}
\mathrm{IMP} = \frac{1}{|\mathcal{P}|}\sum_{p\in\mathcal{P}} \mathrm{IMP}_p,
\end{equation}
where $\mathcal{P}$ is the test pocket set.

{Mean Percentage Binding Gain (MPBG).}
For each pocket $p$, let $E^{\mathrm{ref}}_p$ be the Vina Dock energy of the reference ligand and 
$\{E_{p,j}\}_{j=1}^M$ the Dock energies of $M$ generated ligands.
The mean Dock energy over generated ligands is
\begin{equation}
\bar{E}_p = \frac{1}{M}\sum_{j=1}^M E_{p,j}.
\end{equation}
The per-pocket percentage binding gain is summarized by the Mean Percentage Binding Gain (MPBG),
\begin{equation}
\mathrm{MPBG}
= \frac{1}{\lvert\mathcal{P}\rvert}
  \sum_{p\in\mathcal{P}}
  \frac{E^{\mathrm{ref}}_p - \bar{E}_p}{\lvert E^{\mathrm{ref}}_p\rvert}
  \times 100\%,
\end{equation}
where $\mathcal{P}$ denotes the test pocket set.

{Ligand Binding Efficiency (LBE)}
For a generated ligand with Vina energy $E$ and heavy atom count $N_{\mathrm{HA}}$, its binding efficiency is measured as $E / N_{\mathrm{HA}}$. The model-level LBE is defined as the average binding efficiency over all generated ligands and all test pockets,
\begin{equation}
\mathrm{LBE}
= \frac{1}{\lvert\mathcal{P}\rvert} \frac{1}{M}
  \sum_{p\in\mathcal{P}} \sum_{j=1}^M
  \frac{E_{p,j}}{N_{\mathrm{HA},p,j}},
\end{equation}
where $\mathcal{P}$ denotes the test pocket set and $M$ is the number of generated ligands per-pocket.

\begin{figure*}[htbp] 
\centering \includegraphics[width=0.86\textwidth]{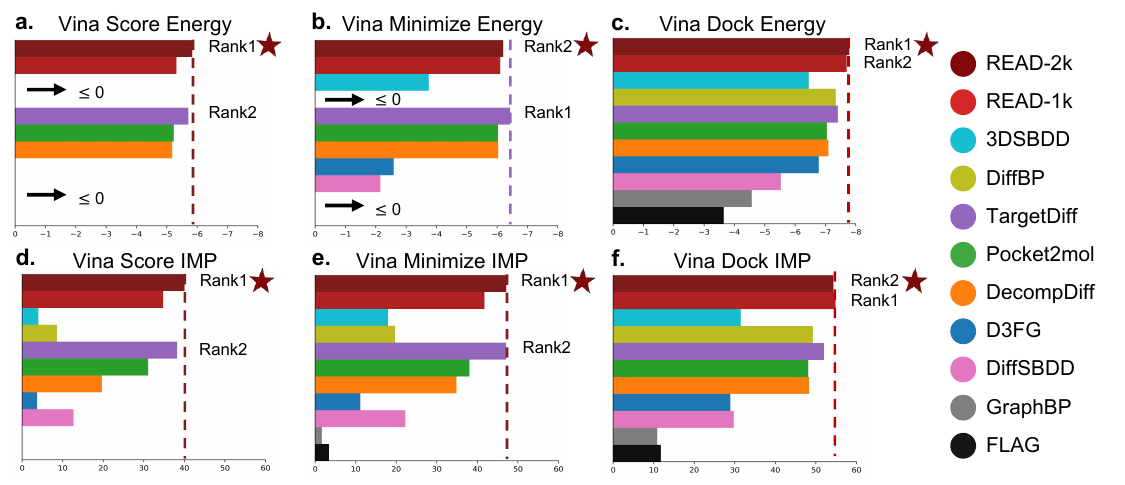} 
\caption{\textbf{Evaluation of READ for pocket specific ligand generation.} The figure compares READ-2k and READ-1k with nine baseline models. \textbf{a--c,} Docking-based binding scores evaluated by three AutoDock Vina metrics. Lower energy (kcal mol$^{-1}$) indicates more favorable Vina scores. READ-2k and READ-1k obtain the two lowest final Vina docking energies in panel (c), which reflects the binding potential after flexible docking. \textbf{d–f,} Improvement Percentage (IMP), defined as the proportion of generated molecules with better Vina scores than the reference ligand (higher is better). READ models generate a larger fraction of improved candidates, especially for Vina IMP after docking in panel (f).} 
\vspace{-4mm}
\label{fig:fig3} 
\end{figure*}

\subsection{Baselines and Configuration}
\label{sec:baselines}
To benchmark READ, comparisons are carried out against nine molecular generators under a unified evaluation protocol. The baselines span three categories: \textbf{autoregressive models} (\textit{3DSBDD}~\cite{3DSBDD}, \textit{GraphBP}~\cite{graphbp}, \textit{Pocket2Mol}~\cite{pocket2mol}), \textbf{diffusion models} (\textit{TargetDiff}~\cite{targetdiff}, \textit{DiffSBDD}~\cite{diffsbdd}, \textit{DiffBP}~\cite{diffbp}), and \textbf{prior guided approaches} (\textit{FLAG}~\cite{flag}, \textit{D3FG}~\cite{d3fg}, \textit{DecompDiff}~\cite{decompdiff}). READ is evaluated in two primary settings using an EGNN backbone~\cite{EGNN}: \textbf{READ-1k} with $T=1000$ diffusion steps and \textbf{READ-2k} with $T=2000$ diffusion steps. Ablation studies are also conducted on a \textit{TargetDiff} backbone without retrieval and alignment modules to isolate their contributions.

\section{Results}\label{sec:results}

This section provides a systematic assessment of the READ paradigm. All methods are evaluated on the CrossDocked2020 benchmark under a unified data split and evaluation pipeline. READ and a subset of baselines are trained under matched configurations. For the remaining baselines, the reported results from CBGBench~\cite{lin2024cbgbench} are adopted. These results were obtained under the same dataset partition and docking based evaluation settings. The impact of READ on docking based binding energy is examined and its performance is compared with nine recent strong molecular generators. In addition to distribution level comparisons based on 100 generated ligands per-pocket, a pocket centric ``one molecule per-pocket'' analysis is performed. For each pocket, the top scoring drug-like ligand among 100 generated candidates that satisfies drug-likeness filters is selected and compared with the native ligand.

\subsection{Docking-Based Ligand--Protein Binding Scores}
\label{sec:binding_energy}

Docking-based ligand--protein binding scores were evaluated with three metrics derived from AutoDock Vina, as defined in Section~\ref{sec:metrics} and summarized in Figures~\ref{fig:fig3}(a--c). The primary metric was the final Vina Dock Energy after flexible docking, complemented by the initial Vina Score Energy before optimization and the Vina Minimize Energy after rigid-body pose refinement. These metrics are used as standardized computational proxies for binding quality rather than as direct measurements of experimental binding affinity.

The Vina Score Energy (Figure~\ref{fig:fig3}(a)) reflects how well a model generates ligands directly in a favorable pose relative to the pocket. This task requires simultaneous design of molecular structure and spatial orientation. Six of the nine baselines yielded Vina Score energies $\ge 0$~kcal~mol$^{-1}$, indicating limited end-to-end pose generation performance. In contrast, READ-2k attained the lowest energy at $-5.84$~kcal~mol$^{-1}$, and READ-1k achieved $-5.32$~kcal~mol$^{-1}$. These results show that both variants recover reasonable pocket–ligand poses without additional pose prediction modules.

The Vina Minimize Energy (Figure~\ref{fig:fig3}(b)) is obtained after rigid body optimization of each generated ligand. READ-2k and READ-1k reached $-6.20$ and $-6.10$~kcal~mol$^{-1}$, respectively. READ-2k ranked second among all methods, slightly behind TargetDiff at $-6.43$~kcal~mol$^{-1}$. The modest change from the Vina Score Energy suggests that READ already places ligands close to favorable poses and requires only limited rigid body adjustment.

The Vina Dock Energy (Figure~\ref{fig:fig3}(c)) measures the most favorable Vina-based docking score after full flexible docking. Under this metric, READ-2k and READ-1k achieved the two strongest results, with energies of $-7.79$ and $-7.70$~kcal~mol$^{-1}$, respectively. Across the three Vina stages under identical evaluation settings, READ-2k provides the most consistent improvement in docking-based binding scores among the compared methods.

\subsection{Improvement Rate Over Reference Ligands}

To assess the overall quality of the generated molecular ensembles, the Improvement Percentage (IMP) metric~\cite{diffbp} is used as defined in Section~\ref{sec:metrics}. For each test pocket and each energy perspective (Score, Minimize, or Dock), the proportion of generated molecules whose energy is lower than that of the reference ligand is computed. The IMP for a given energy metric is obtained by averaging these proportions across all test pockets. A higher IMP therefore indicates that a larger fraction of generated molecules outperform the reference ligand.

\begin{figure}[t] 
\centering\includegraphics[width=0.50\textwidth]{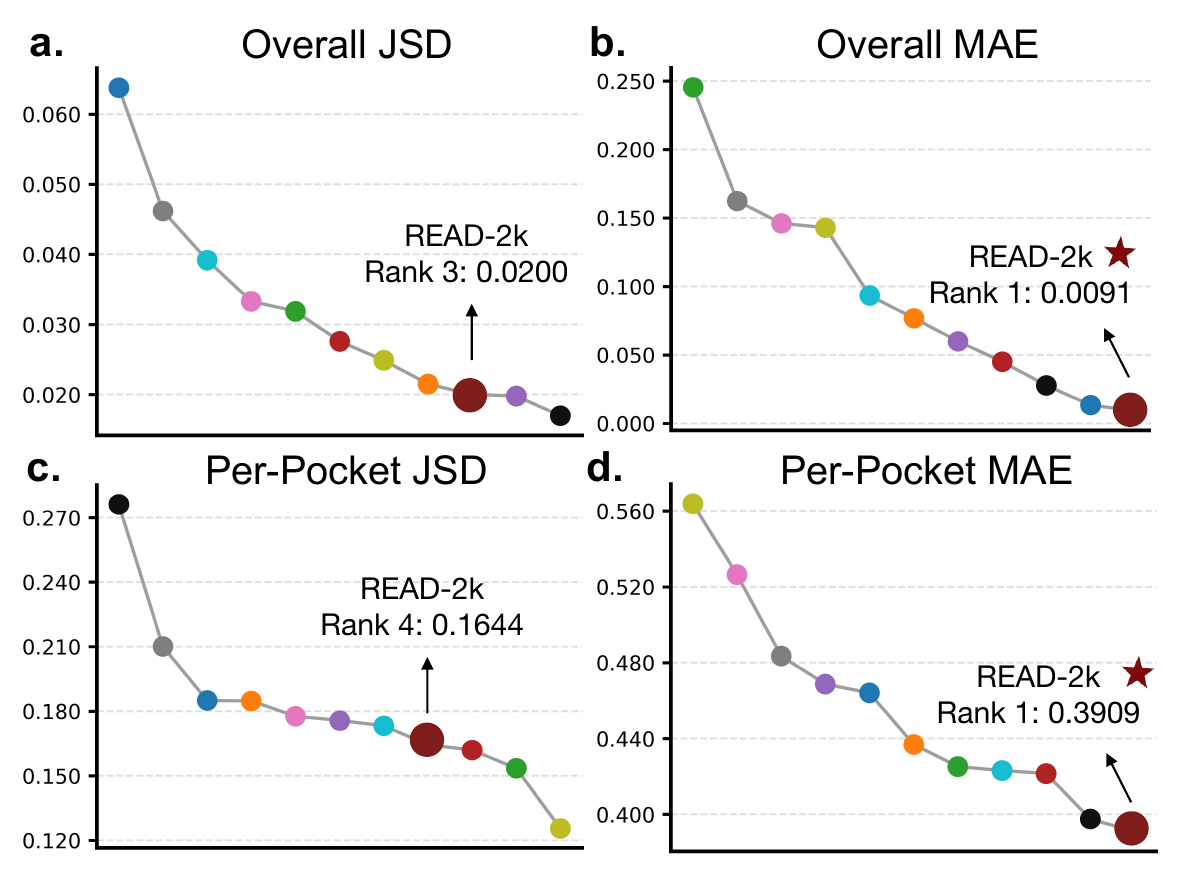} 
\caption{\textbf{Analysis of ligand–protein interaction profiles using four metrics.} Lower values indicate a closer match to the reference interaction patterns. Panels (a) and (c) report Jensen–Shannon divergence (JSD) between interaction type distributions, evaluated globally over all ligands and as a per-pocket average, respectively. Panels (b) and (d) report mean absolute error (MAE) in the counts of interactions per ligand, again evaluated globally and per-pocket. READ-2k ranks among the top methods on both JSD and MAE, indicating that it models both interaction types and interaction counts accurately with respect to the reference complexes.}
\vspace{-4mm}
\label{fig:fig4} 
\end{figure}

\begin{figure*}[htbp] 
\centering \includegraphics[width=0.90\textwidth]{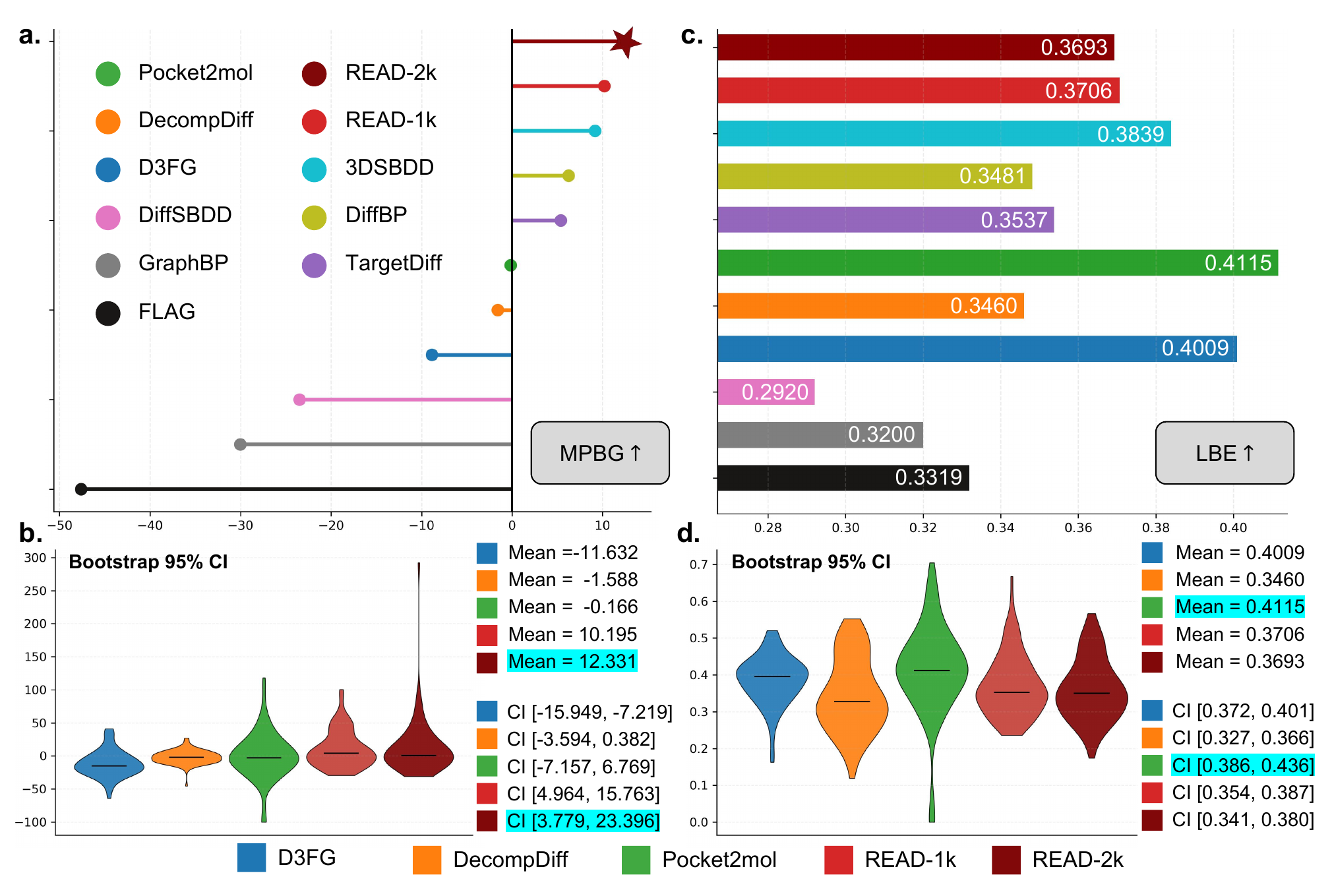}
\vspace{-4mm}
\caption{\textbf{Stability and efficiency of pocket specific generation.} MPBG and LBE over the $100$ test pockets. Panels (a,c) report mean values. Panels (b,d) report per-pocket distributions with bootstrap $95\%$ confidence intervals.}
\label{fig:fig5} 
\end{figure*}

The Vina Score IMP, which evaluates the initial unoptimized poses (Figure~\ref{fig:fig3}(d)), reveals a key strength of READ. In this challenging setting, READ-2k and READ-1k produce a higher proportion of superior candidates, with IMP values of $40.04\%$ and $34.75\%$, respectively. In contrast, the baseline models show substantially weaker performance. This behavior is consistent with known limitations of autoregressive models, whose sequential atom placement makes it difficult to maintain global pocket–ligand interactions. In contrast, the global denoising process of diffusion models is inherently better suited to modeling the coupled energy of an entire molecule.

After rigid body pose optimization (Vina Minimize IMP, Figure~\ref{fig:fig3}(e)), the READ models perform on par with the strongest baseline, TargetDiff. The strong performance of TargetDiff can be attributed to its use of an atom count prior based on pocket scale that is explicitly encoded in the model. READ achieves a comparable outcome without an additional loss term by leveraging retrieval and alignment. These components implicitly establish a mapping between pocket scale and atomic composition.

The most pronounced improvements emerge after the final flexible docking stage (Vina Dock IMP, Figure~\ref{fig:fig3}(f)). At this point, READ-1k and READ-2k achieve success rates of $54.82\%$ and $54.31\%$, respectively. These values establish a substantial advantage within the docking-based evaluation protocol. This result highlights distinct limitations of alternative approaches. Grid based methods often fail to resolve fine grained poses in complex pockets. Models with strong explicit priors, such as DecompDiff, can overly restrict conformational exploration. Centroid based models such as DiffBP depend on an accurate initial placement, and their performance is compromised when the centroid prediction is suboptimal.

In summary, the READ paradigm consistently increases the proportion of candidates with favorable docking-based scores across all three evaluation perspectives. Its advantage in the initial Score phase demonstrates effective generation of molecules with appropriate shapes in the correct spatial regions. This advantage is maintained after rigid optimization and is further strengthened after flexible docking. These observations confirm that READ achieves a robust match between interaction patterns and pocket constraints. In contrast, methods that rely on decompositional priors or separate pre-generation modules show limited ability to sustain such improvements across all energy metrics.

\subsection{Ligand-Protein Interaction Profiles}

To obtain a detailed view of the generated ligand–protein interactions, PLIP was used to categorize contacts into seven types: hydrogen bonds, hydrophobic interactions, salt bridges, $\pi$–stacking, $\pi$–cation, metal coordination, and water bridges. Four metrics were defined to compare the interaction profiles of generated ligands with those of the reference ligands. Overall JSD and per-pocket JSD use the Jensen–Shannon divergence (JSD) to measure the similarity of interaction-type distributions, either aggregated over all samples or averaged over pockets. Overall MAE and per-pocket MAE use the mean absolute error (MAE) to quantify the difference in the average number of interactions per ligand, again evaluated globally and on a per-pocket average.

The JSD metrics characterize how well each model reproduces the relative proportions of different interaction types. In the global comparison (overall JSD, Figure~\ref{fig:fig4}(a)), READ-2k attained the third-best score (0.0200). This behavior is consistent with the fact that this metric strongly favors methods with explicit priors, such as FLAG and DecompDiff, which calibrate interactions toward frequent patterns in the training data~\cite{flag,decompdiff}, as well as diffusion models like TargetDiff that effectively restore global distributions. When evaluated on a per-pocket basis (per-pocket JSD, Figure~\ref{fig:fig4}(c)), READ again ranks near the top (third and fourth place). This setting slightly favors models with strong local constraints, such as autoregressive methods (Pocket2Mol) or approaches that predefine a focused generation region (DiffBP)~\cite{pocket2mol,diffbp}.

In contrast, the MAE metrics assess whether the absolute number of interactions per ligand is matched correctly. On these metrics, the advantage of READ becomes clear (Figure~\ref{fig:fig4}(b,d)). READ-2k attains the best performance in both overall MAE (0.0091) and per-pocket MAE (0.3909), indicating that the retrieval and alignment mechanism keeps the quantity and strength of key interactions within the reference range, while many baselines either overproduce or underproduce contacts for individual pockets.

Taken together, the four metrics show that READ is the only method that maintains strong performance across both distributional fidelity (JSD) and interaction counts (MAE), whereas other baselines exhibit pronounced trade-offs between these aspects. This indicates that READ combines the strengths of explicit-prior methods and locally constrained models: it represents global and local interaction patterns accurately and remains stable across diverse pocket geometries, rather than relying on a single type of prior or constraint.

\begin{figure*}[!t] 
\centering \includegraphics[width=0.99\textwidth]{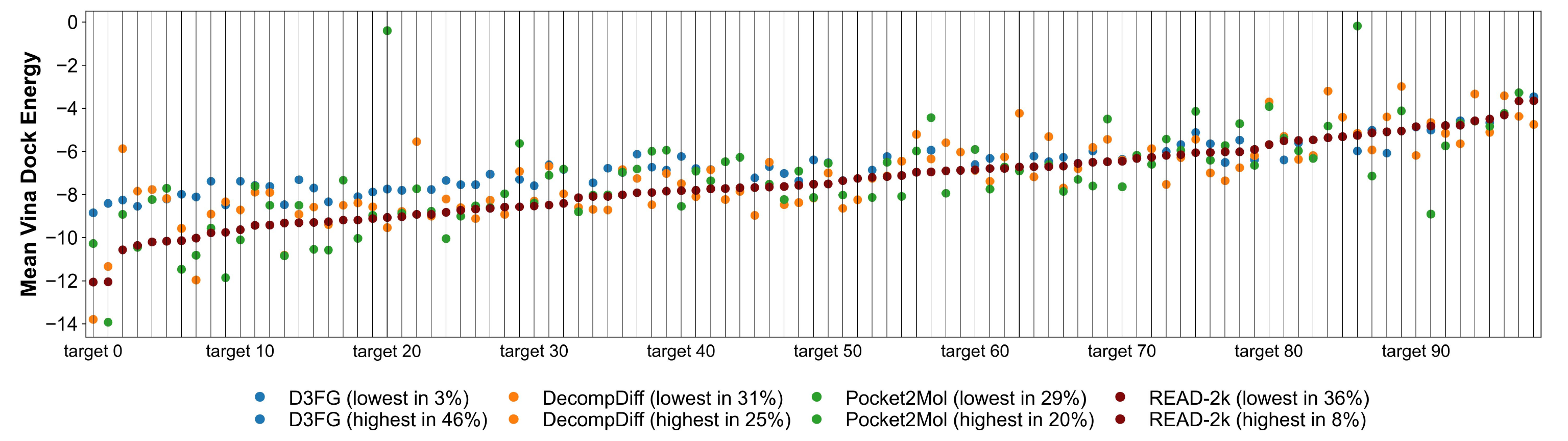} 
\vspace{-4mm}
\caption{\textbf{Per-pocket performance comparison of the mean Vina Dock Energy for the top five methods.} Each point represents the average energy for a single target pocket. The plot illustrates that READ-2k not only frequently achieves the best performance (lowest energy, indicated as ``highest'' in rank for 36\% of targets) but also rarely fails, having the lowest rank in only 8\% of cases. This showcases its high performance floor and reliability.}
\vspace{-2mm}
\label{fig:fig6} 
\end{figure*}

\subsection{Stability of Pocket-Specific Generation Quality}\label{sec:stability}

Generation quality is expected to be stable across targets rather than driven by a few easy cases. Although the CrossDocked2020 test set contains $100$ pockets, this split follows the standard benchmark setting used in prior SBDD studies and is adopted here to ensure direct comparison under a unified protocol. To quantify stability under this benchmark, two complementary metrics are considered. Mean Percentage Binding Gain (MPBG) measures, for each pocket, the percentage change in the mean Vina Dock energy of the generated ligands relative to the reference ligand, and then averages this quantity over all pockets. Ligand Binding Efficiency (LBE)~\cite{lbeused} is defined as the docking energy normalized by the number of heavy atoms, which controls for trivial gains obtained by generating very large ligands.

Figure~\ref{fig:fig5} summarizes MPBG and LBE over the $100$ test pockets: panels (a,c) report mean values, and panels (b,d) show per-pocket distributions with bootstrap $95\%$ confidence intervals.
Figures~\ref{fig:fig5}(a,b) show that READ is the only family of methods with consistently positive MPBG. The mean MPBG of READ-2k reaches $12.33\%$, with a bootstrap $95\%$ confidence interval $[3.78, 23.40]$, and READ-1k attains $10.20\%$ with a confidence interval $[4.96, 15.76]$. In contrast, diffusion and prior guided baselines such as D3FG and DecompDiff exhibit negative or near zero mean MPBG, and their confidence intervals either lie entirely below zero or straddle it. These results indicate that READ improves Vina-based docking scores for most targets, whereas several baselines on average reduce docking-based scores relative to the reference ligand.

The LBE statistics in Figures~\ref{fig:fig5}(c,d) provide a complementary view. Pocket2Mol attains the highest mean LBE by generating particularly compact ligands, but its variance is large. READ-1k and READ-2k achieve slightly lower mean LBE, yet their confidence intervals are noticeably narrower and remain competitive with the best baseline. This pattern suggests that READ attains a more reliable balance between binding energy and ligand size: binding improvements are obtained without relying on extreme miniaturization or large pocket-specific fluctuations.

Per-pocket rank statistics further confirm this stability advantage. Figure~\ref{fig:fig6} reports, for the five strongest methods, the distribution of ranks in terms of mean Vina Dock energy over pockets. READ-2k attains the best energy on $36\%$ of pockets and ranks last on only $8\%$ of pockets, whereas competing methods are both less frequently best and more often worst. Together with its top MPBG and positive confidence interval, this rank profile indicates that READ raises the performance floor across targets; improvements are widespread rather than driven by a small number of favorable pockets.

\subsection{Component-Level Ablation}
\label{sec:component_ablation}

\begin{figure}[htbp] 
\centering \includegraphics[width=0.40\textwidth]{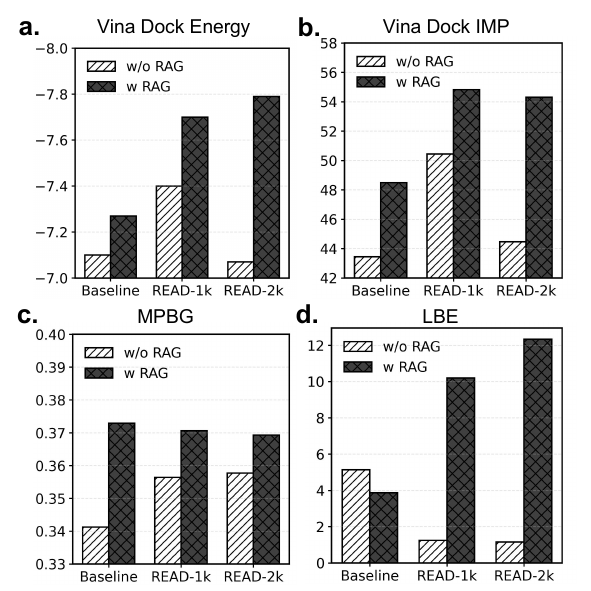}
\vspace{-4mm}
\caption{\textbf{a--d,} Ablation study on the effect of retrieval-augmented conditioning. The comparison between models without retrieval augmentation (w/o RAG) and models with retrieval augmentation (w/ RAG) is shown for Vina Dock Energy (a), Vina Dock IMP (b), MPBG (c), and LBE (d). A more detailed component-level ablation separating retrieval, inverted alignment, and inference-time guidance is provided in Supplementary Table S1.}
\label{fig:fig7} 
\end{figure}

The contribution of retrieval and alignment was first examined in Figure~\ref{fig:fig7}. 
On the plain diffusion backbone (\emph{Baseline}), adding retrieval-augmented conditioning without inverted layer-wise alignment (\emph{Baseline w/ RAG}) leads to only modest and sometimes inconsistent changes across the four metrics. 
In contrast, when retrieval is combined with the alignment-adapted denoiser in READ-1k and READ-2k, the improvements in Vina Dock Energy, Vina Dock IMP, MPBG, and LBE become more consistent.

To further disentangle the roles of retrieval, inverted layer-wise alignment, and inference-time guidance, we provide a component-level ablation in Supplementary Table~S1. 
The results show that retrieval alone is not sufficient to explain the performance gain, whereas coupling retrieval with representation alignment and scheduled guidance yields stable improvements in docking-oriented metrics. 
These observations support the align-then-guide design of READ.

\begin{table*}[!t]
\centering
\caption{Substructure fidelity, static geometry, and clash ratios.}
\label{tab:substructure_geometry_clash}
\setlength{\tabcolsep}{4.2pt}      
\renewcommand{\arraystretch}{1.25} 
\begin{tabular}{llcccccccccc}
\toprule
\multirow{2}{*}{Group} & \multirow{2}{*}{Method} &
\multicolumn{2}{c}{Atom type} &
\multicolumn{2}{c}{Ring type} &
\multicolumn{2}{c}{Functional group} &
\multicolumn{2}{c}{Static Geometry} &
\multicolumn{2}{c}{Clash} \\
& & JSD $\downarrow$ & MAE $\downarrow$ & JSD $\downarrow$ & MAE $\downarrow$ &
JSD $\downarrow$ & MAE $\downarrow$ & JSD$_{\mathrm{BL}}\downarrow$ & JSD$_{\mathrm{BA}}\downarrow$ &
Ratio$_{\mathrm{cca}}\downarrow$ & Ratio$_{\mathrm{cm}}\downarrow$ \\
\midrule
\multirow{3}{*}{Autoregressive}
& 3DSBDD~\cite{3DSBDD}     & 0.0860 & 0.8444 & 0.3188 & 0.2457 & 0.2682 & 0.0494 & 0.5024 & 0.3904 & 0.2482 & 0.8683 \\
& GraphBP~\cite{graphbp}   & 0.1642 & 1.2266 & 0.5061 & 0.4382 & 0.6259 & 0.0705 & 0.5182 & 0.5645 & 0.8634 & 0.9974 \\
& Pocket2Mol~\cite{pocket2mol} & 0.0916 & 1.0497 & 0.3550 & 0.3545 & 0.2961 & 0.0622 & 0.5433 & 0.4922 & 0.0576 & 0.4499 \\
\midrule
\multirow{6}{*}{Diffusion}
& TargetDiff~\cite{targetdiff}  & 0.0533 & 0.2399 & 0.2345 & \underline{0.1559} & 0.2876 & 0.0441 & 0.2659 & \underline{0.3769} & 0.0483 & 0.4920 \\
& DiffBP~\cite{diffbp}          & 0.2591 & 1.5491 & 0.4531 & 0.4068 & 0.5345 & 0.0670 & 0.3453 & 0.4621 & 0.0449 & 0.4077 \\
& DiffSBDD~\cite{diffsbdd}      & 0.0529 & 0.6316 & 0.3853 & 0.3437 & 0.5520 & 0.0710 & 0.3501 & 0.4588 & 0.1083 & 0.6578 \\
& FLAG~\cite{flag}              & 0.1032 & 1.7665 & 0.2432 & 0.3370 & 0.3634 & 0.0666 & 0.4215 & 0.4324 & 0.6777 & 0.9769 \\
& D3FG~\cite{d3fg}              & 0.0644 & 0.8154 & \underline{0.1869} & 0.2204 & \underline{0.2511} & 0.0516 & 0.3727 & 0.4700 & 0.2115 & 0.8571 \\
& DecompDiff~\cite{decompdiff}  & \textbf{0.0431} & 0.3197 & 0.2431 & 0.2006 & \textbf{0.1916} & \textbf{0.0318} & \underline{0.2576} & \textbf{0.3473} & 0.0462 & 0.5248 \\
\midrule
\multirow{3}{*}{Ours}
& READ-1k       & 0.0491 & \underline{0.2096} & 0.2512 & 0.2153 & 0.3246 & 0.0480 & 0.2885 & 0.4625 & 0.0483 & 0.4870 \\
& READ-2k       & \underline{0.0489} & \textbf{0.1728} & 0.2563 & 0.2480 & 0.3166 & 0.0508 & 0.3371 & 0.5140 & \underline{0.0262} & \underline{0.3709} \\
& READ-2k-prior & 0.0537 & 0.6269 & \textbf{0.0815} & \textbf{0.0654} & 0.2789 & \underline{0.0380} & \textbf{0.2170} & 0.4433 & \textbf{0.0255} & \textbf{0.3523} \\
\bottomrule
\end{tabular}

\vspace{1mm}
\begin{minipage}{\textwidth}
\footnotesize \emph{Note:} Lower is better for all metrics ($\downarrow$). \textbf{Bold} indicates the best result in each column; \underline{underline} indicates the second best.
\vspace{-4mm}
\end{minipage}
\end{table*}

\subsection{Substructure Distribution, Static Geometry, and Steric Clashes}

The chemical plausibility of the generated candidates was evaluated along three axes: substructure distributions, static geometry, and steric clashes with the receptor. Substructure quality was measured for atom types, ring types, and functional groups by Jensen–Shannon divergence (JSD) and mean absolute error (MAE) between the generated and reference distributions. Static geometry was characterized by JSD over bond length and bond angle distributions (JSD$_{\mathrm{BL}}$ and JSD$_{\mathrm{BA}}$). Steric quality was summarized by two clash ratios, Ratio$_{\mathrm{cca}}$ and Ratio$_{\mathrm{cm}}$, which denote the fractions of ligand atoms involved in close contacts within the pocket and in severe clashes, respectively. All metrics are reported in Table~\ref{tab:substructure_geometry_clash}, where lower values indicate better agreement with the reference.

The READ models achieved a balanced trade-off between substructure fidelity and geometric correctness. READ-2k attained the lowest atom type MAE among all methods (0.1728) while keeping atom type JSD competitive with the best diffusion baseline (0.0489 versus 0.0431 for DecompDiff). Methods with strong fragment priors, such as D3FG and DecompDiff, obtained the leading JSD for ring types and functional groups (for example a ring type JSD of 0.1869 for D3FG and a functional group JSD of 0.1916 for DecompDiff), but this advantage was accompanied by larger errors in atom type counts and noticeably higher clash ratios. READ-2k, in contrast, maintained low clash ratios and preserved static geometry close to the reference: its bond length and bond angle JSDs were comparable to or better than those of several diffusion baselines that did not use retrieval or alignment. These patterns indicated that alignment guided denoising coordinated local chemistry and global pose so that ligands fitted the pocket without introducing excessive strain or overlap, whereas autoregressive generators tended to accumulate geometric distortions along the generation sequence.

To further probe the effect of structural priors, a variant READ-2k-prior was evaluated with a single BRICS fragment fixed during generation. This light structural anchor improved ring system and geometry statistics: READ-2k-prior achieved the best ring type JSD (0.0815), the best ring type MAE (0.0654), the lowest bond length JSD (0.2170), and the lowest clash ratios in the entire comparison (0.0255 and 0.3523). These gains came at the cost of a higher atom type MAE relative to READ-2k, but the overall profile confirmed that the Retrieval, Alignment, and Diffusion framework already provided competitive substructure and geometry statistics without heavy structural priors, and that its performance could be further enhanced by a minimal fragment anchor. Retrieval and representation alignment supplied an effective inductive bias for spatial–chemical correspondence while retaining sufficient flexibility for adaptation to diverse pocket environments.

\begin{figure*}[htbp] 
\centering \includegraphics[width=0.96\textwidth]{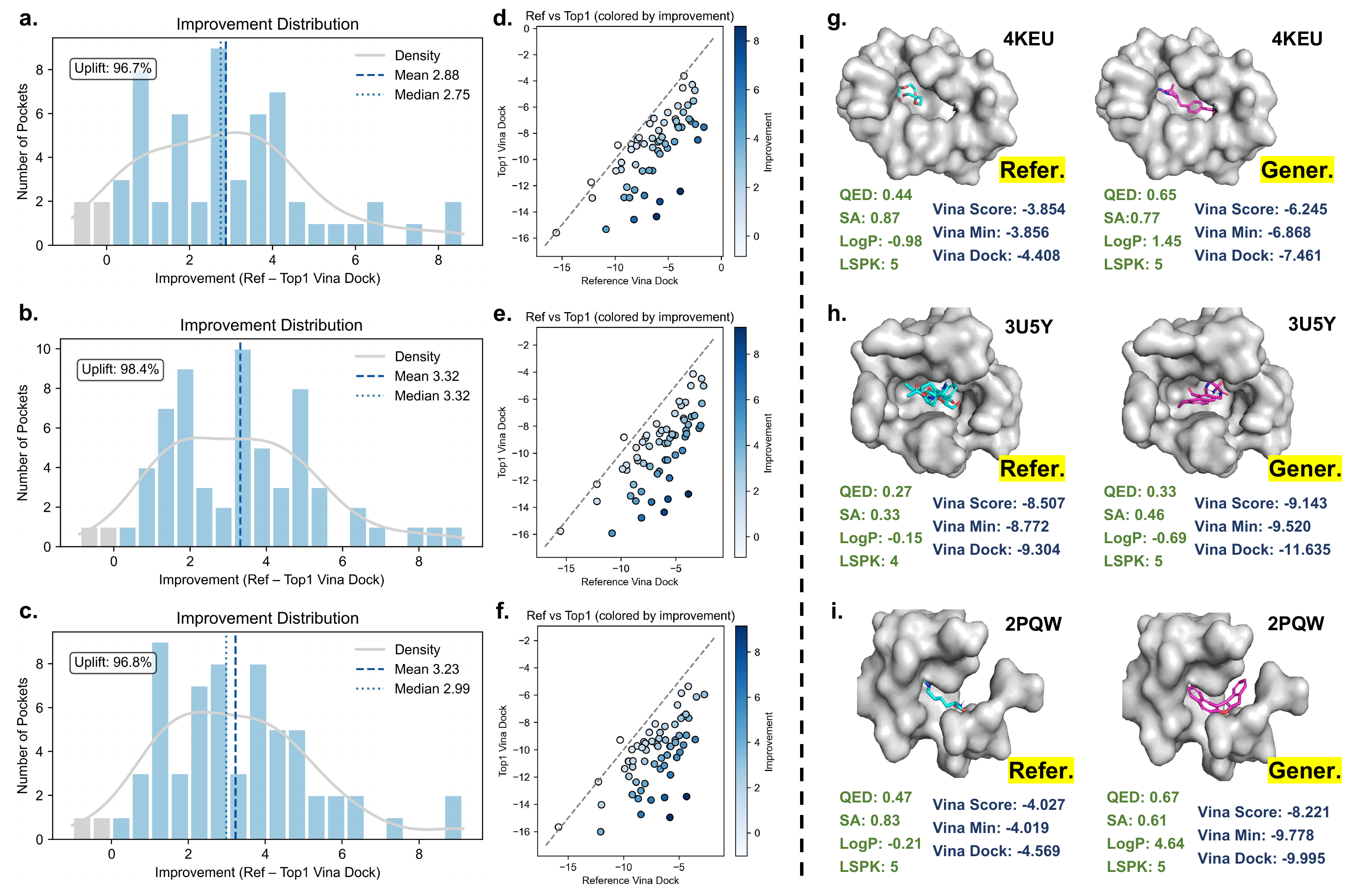}
\vspace{-4mm}
\caption{\textbf{Per-pocket selection analysis and representative qualitative examples.}
Panels (a–c) report per-pocket docking gains under the selection protocol, where for each pocket the top-scoring ligand is selected among $100$ generated candidates after drug-likeness filtering (QED $>0.5$, SA $<4$, and satisfying all five Lipinski rules); gray bars indicate pockets with degraded scores and blue bars indicate improved scores.
Panels (d–f) plot reference versus selected docking metrics for Vina Score, Vina Minimize, and Vina Dock, with the dashed diagonal indicating parity.
Panels (g–i) show three representative pockets (4KEU, 3U5Y, and 2PQW) illustrating typical improvement patterns.}
\label{fig:fig8} 
\end{figure*}

\section{Discussion}
\label{sec:discussion}

\subsection{Relation to Prior Approaches and Main Implications}

Pocket conditioned diffusion and flow models have established a strong baseline for structure based molecular generation by combining learned pocket encoders, joint coordinate and type variables, and SE(3) equivariant architectures~\cite{targetdiff,diffsbdd,unimomo,paflow}. Other lines of work inject domain knowledge by physics guided losses~\cite{irdiff,pidiff}, pharmacophore or fragment priors~\cite{d3fg,PharmaDiff,flag,fraggen,resgen}, or retrieval mechanisms.

READ differs from prior approaches along three main aspects. First, it is pocket centric. Retrieval is driven by pocket homology rather than ligand similarity, so the inductive bias is attached to the target environment instead of to a fixed library of molecules. Second, it separates the construction of priors from their use. Pocket information is translated once into ligand side shape, interaction, and atom preference signals. These signals are then aligned with the denoiser across depth, rather than being injected as raw protein features at every layer. Third, it couples training time alignment with inference time guidance. The model is first trained so that denoiser states and encoder representations share a common geometry. Time scheduled gradients from priors are then applied along the diffusion trajectory at inference.

Empirically, this combination had two implications. Binding energy and improvement metrics were raised without degrading interaction profile fidelity, static geometry, or substructure statistics. This suggests that retrieval does not merely overfit to common motifs. In addition, per-pocket metrics such as mean percentage binding gain and ligand binding efficiency showed that gains were spread across many pockets. These metrics also indicated that variance was reduced relative to several strong baselines. Taken together, these observations indicate that nonparametric priors drawn from a protein–ligand atlas can act as a stabilizing bias for parametric diffusion models. This effect relies on the mapping between atlas features and denoiser states being explicitly aligned.

\subsection{Benchmark, Docking Metrics, and Selection Protocol}
\label{sec:limitations}

The conclusions in this work are constrained by the properties of the CrossDocked2020 benchmark and the docking-based evaluation protocol. The $100$-pocket test set follows the standard split used in prior structure-based molecular generation studies, which enables direct comparison with existing methods but also limits the statistical scope of the benchmark. CrossDocked2020 is derived from experimentally resolved complexes and cross docking, but it remains a static snapshot of protein conformations and drug-like chemistry. Certain target families and ligand chemotypes are more heavily represented than others, and pockets that undergo large induced fit or allosteric changes are underrepresented. In such cases, the retrieval atlas is also biased toward protein families and chemotypes that appear in available structures. When a query pocket lies in a sparsely populated region of this atlas, retrieval returns more distant neighbors. In that case, the effective guidance from priors becomes weaker. In the extreme, READ is expected to behave more like its underlying diffusion backbone, with limited benefit from retrieval. We further characterize retrieval similarity and atlas coverage in Supplementary Table S4 to clarify how the strength of retrieved homologous priors relates to pocket-level performance.

The reliance on docking scores introduces further limitations. AutoDock Vina and related scoring functions are standardized and widely used for SBDD benchmarking, but they remain computational proxies rather than direct experimental measurements of binding affinity. They may approximate physical binding free energies imperfectly and can favor larger ligands that form many contacts. In this work, binding energy metrics were complemented by ligand binding efficiency, which normalizes docking energy by heavy atom count. This was intended to reduce the bias toward very large molecules. READ achieved strong performance on both energy and efficiency metrics, but these remain in silico proxies. Experimental affinity measurements and prospective validation would be needed to fully assess how gains in docking space translate into gains in true binding affinity.

Overall, the benchmark and scoring choices were sufficient to reveal systematic differences between READ and strong baselines under a controlled setting, but they also limit the scope of the claims. Extending the analysis to other datasets, alternative scoring functions, and experimental readouts remains an important direction.

\subsection{Per-Pocket Behavior and Qualitative Patterns}
\label{sec:case_studies}

Generation quality is expected to be stable across targets rather than driven by a few easy cases. The per-pocket “one molecule per-pocket” analysis was therefore designed to approximate a simple hit triage procedure. For each pocket, candidates were first filtered by drug-likeness criteria (QED greater than $0.5$, synthetic accessibility below a threshold, and compliance with Lipinski’s rule of five). The top scoring ligand was then selected from the remaining candidates. This introduces a bias toward chemically attractive and synthetically plausible molecules. However, it is consistent with common practice in computational chemistry, where docking scores are interpreted only after basic medicinal chemistry filters have been applied. Under this protocol, the results should be interpreted as showing that READ increases the probability of finding candidates that simultaneously satisfy standard drug-likeness filters and exhibit improved docking scores. They do not directly address novelty, scaffold diversity, or late stage developability.

Under this protocol, the improvement probability in Fig.~\ref{fig:fig8}(a–c) is computed as the fraction of pockets whose selected ligand improves upon the native ligand under Vina Score, Vina Minimize, and Vina Dock, respectively, and the corresponding mean score reductions are reported in the same panels.
Figure~\ref{fig:fig8}(d–f) further plots, for each pocket, the selected docking metric against the reference metric under the three Vina stages, where points below the diagonal indicate improvements and the color encodes the magnitude of the gain.
Within this setting, per-pocket statistics show that READ improved docking-based scores for a substantial fraction of pockets. READ also rarely produced the worst result among the strongest methods. Qualitative inspection of generated complexes suggests that many gains are associated with recurring patterns in how pockets are exploited. Three representative cases illustrate these mechanisms (Figure~\ref{fig:fig8}).

In pocket 4KEU, the reference ligand left a central void. The READ generated ligand extended along the main channel into a side subpocket and occupied previously unused volume. Meanwhile, it reduced the number of rotatable bonds. The resulting pose formed an additional hydrogen bond network with residues such as TYR97, ASP256, and THR265. It also increased the number of hydrophobic contacts. In pocket 3U5Y, which contains a dense set of polar residues, the READ ligand formed additional directional hydrogen bonds with GLN186, HIS140, and TRP477. It positioned its aromatic core close to the pocket surface and satisfied polar contacts that were only partially addressed by the reference ligand. In pocket 2PQW, the reference ligand occupied one side of a narrow hydrophobic groove. The READ ligand used a multicyclic scaffold to span both sides of this region, maintained a key hydrogen bond at the base of the groove, and increased the extent of hydrophobic contacts.

These examples illustrate typical ways in which retrieval and alignment can correct common deficiencies of reference ligands, such as empty subpockets, unsatisfied polar sites, or fragmented hydrophobic coverage. However, the method did not improve all pockets equally. Pockets with already optimized reference ligands or with limited atlas coverage showed only marginal gains. The per-pocket rank distributions and confidence intervals therefore should be viewed as characterizing substantial variation across pockets. READ improves both the average performance and the lower tail of the distribution, but further gains are still possible on challenging or underrepresented targets.

\subsection{Computational Cost and Scalability}
\label{sec:cost_scalability}

The main computational cost of READ comes from diffusion sampling, which scales approximately linearly with the number of denoising steps and the number of generated candidates. The retrieval step is performed once per target pocket and can be cached, so its overhead is separated from iterative sampling. READ-1k provides a more cost-efficient setting and achieves performance close to READ-2k, whereas READ-2k is useful when maximizing docking-space performance is preferred over generation speed. Detailed training and inference costs, including encoder pretraining, denoiser training, retrieval overhead, memory usage, and generation time per pocket, are reported in Supplementary Table S3.

\section{CONCLUSION}

This work presented READ, a retrieval alignment diffusion framework for structure based molecular generation. For each target pocket, ligands from homologous proteins were retrieved and distilled into ligand side priors for shape and volume occupancy, interaction patterns, and atom and bond preferences. These priors were aligned with an equivariant denoiser across network depth and were introduced along a time scheduled diffusion process. On the CrossDocked2020 benchmark under a unified docking and interaction analysis protocol, READ achieved strong performance on docking based binding energies and improvement rates over reference ligands. Meanwhile, it maintains competitive interaction profile fidelity, substructure distributions, static geometry, and clash ratios. Pocket-specific metrics such as mean percentage binding gain and ligand binding efficiency indicated that retrieval and representation alignment jointly improved both the average docking-based quality and the stability of pocket-specific generation. Extensions to flexible receptor modeling, richer biochemical priors, and experimental feedback would help clarify how gains in docking space translate into improvements in experimentally confirmed affinity.

\section{Acknowledgments}
This work was supported by the National Natural Science Foundation of China under Grant 6247617, the Guangdong Natural Science Foundation Project under Grant 2025A1515011567. This work is supported by the Intelligent Computing Center of Shenzhen University.

 

\bibliographystyle{IEEEtran}
\bibliography{IEEEabrv,bibliography}

\end{document}